%% file: AMLC2.tex
\documentclass[journal, 10pt]{IEEEtran}
\usepackage{graphics}
\usepackage{epsfig}

\usepackage{subfigure}
\usepackage{graphics}

\usepackage{amsmath}
\usepackage{amsfonts}
\usepackage{amssymb}
\usepackage{stfloats}

\input{shortcuts.tex}

\rem{

\renewcommand{\P}{{\rm P}}

}

\normalsize
\begin{document}

\title{The approximate maximum-likelihood certificate \thanks{This research was supported by the Israel Science Foundation, grant no. 772/09.}}

\author{\begin{tabular}{c c} Idan~Goldenberg,~\IEEEmembership{Student Member,~IEEE} & David~Burshtein,~\IEEEmembership{Senior Member,~IEEE} \\
School of Electrical Engineering & School of Electrical Engineering \\ Tel-Aviv University, Tel-Aviv 69978, Israel & Tel-Aviv University, Tel-Aviv 69978, Israel\\
Email:  idang@eng.tau.ac.il & Email:  burstyn@eng.tau.ac.il
\end{tabular} }


\maketitle 

\IEEEpeerreviewmaketitle

\begin{abstract}
A new property which relies on the linear programming (LP) decoder,
the \emph{approximate maximum-likelihood certificate} (AMLC), is
introduced. When using the belief propagation decoder, this property
is a measure of how close the decoded codeword is to the LP
solution. Using upper bounding techniques, it is demonstrated that
the conditional frame error probability given that the AMLC holds
is, with some degree of confidence, below a threshold. In channels
with low noise, this threshold is several orders of magnitude lower
than the simulated frame error rate, and our bound holds with very
high degree of confidence. In contrast, showing this error
performance by simulation would require very long Monte Carlo runs.
When the AMLC holds, our approach thus provides the decoder with
extra \emph{error detection} capability, which is especially
important in applications requiring high data integrity.
\end{abstract}


\section{Introduction}
Linear programming (LP) decoding has emerged in recent years as a
potential candidate algorithm for approximating maximum-likelihood
(ML) decoding. One reason for this is that it has been shown
\cite{lpdecode} that the LP decoding algorithm has the \emph{ML
certificate} property, i.e., that if the decoder outputs a valid
codeword, it is guaranteed to be the ML codeword.

Since the discovery of LP decoding, several papers have been written
on the subject of improving the performance of the decoder, e.g., by
using integer programming or mixed integer linear
programming~\cite{lpdecode},~\cite{draper2007mdv}, adding
constraints to the Tanner graph~\cite{taghavi2008amlp} and guessing
facets of the polytope~\cite{dimakis2007gfp}. Moreover, the issue of
decoding complexity has been addressed
\cite{low_complexity_LP},~\cite{low_complexity_LP_journal},~\cite{yang2008nlp},
as the complexity of linear programming techniques is in general
polynomial but not linear in the block length $N$. Vontobel and
Koetter~\cite{low_complexity_LP} have proposed an iterative,
Gauss-Seidel-type algorithm for approximate LP decoding. Based on
their general approach, a linear-complexity ($O(N)$) iterative
approximate decoder~\cite{lpon_journal} was suggested.

This low-complexity LP decoder was recently put to use in a
framework \cite{burshtein2010mllp} aimed at harnessing the LP
decoder for tasks other than decoding. In this context, an algorithm
with complexity $O(N^2)$ was proposed which produces a lower bound
on the minimum distance of a specific code. Another use is an
algorithm of the same complexity for finding a tight lower bound on
the fractional distance.

In this paper we propose a new application for using the LP decoder
by introducing a new concept, the \emph{approximate ML certificate}
(AMLC), a tool which can improve the error detection capability of
the belief propagation (BP) decoder. We show that if the BP decoder
output satisfies the AMLC property (in particular, it is a
codeword), then there is a high degree of certainty that it is the
correct codeword. It is demonstrated that, when applying this
technique within the error floor region, the frame error rate
implied by the AMLC is several orders of magnitude lower than the
average rate (the average rate in the error floor region was
previously studied by Richardson \cite{richardson2003efl});
ascertaining this improved reliability directly using Monte Carlo
simulation would require very long simulation runs. This makes the
AMLC especially useful in applications where a high level of data
integrity is required.

The LP decoder is a central component in the evaluation of the AMLC,
as is the aforementioned $O(N^2)$ algorithm for obtaining a lower
bound on the minimum distance. Another component used in our
analysis is the generalized second version of the Duman-Salehi
bound, as derived by Sason and Shamai \cite{shamai2002vgb}; this
bound is an upper bound on the ML decoding probability. A slightly
modified version of this bound is used for our purposes.

This paper is organized as follows. Section~\ref{sec: preliminaries}
provides some background material, related primarily to the LP
decoder. In section~\ref{sec:approx_ML_cert}, we prove our main
result concerning the approximate ML certificate. In
Section~\ref{sec: numerical results} numerical examples are
provided, and the paper is concluded in Section~\ref{sec:
conclusion}.

\section{Preliminaries}\label{sec: preliminaries}
Consider an LDPC code $\cC$ described by a Tanner graph with $N$
variable nodes and $M$ check nodes. A codeword $\bc\in \cC$ is
transmitted over a discrete memoryless binary-input,
output-symmetric (MBIOS) channel described by a probability
transition function $Q(y \given c)$, where $c$ is the transmitted
code-bit and $y$ is the channel output (we will also use the vector
notation $Q(\by \given \bc)$ where $\by$ is the channel output
vector and $\bc$ is the transmitted codeword, the meaning will be
clear from the context). Following the notation in~\cite{lpdecode},
let $\cI$ and $\cJ$ be the sets of variable and constraint nodes,
respectively, such that $|\cI| = N$ and $|\cJ|=M$. Define the set
$\cN_i$ to be the set of neighbors of variable node $i\in\cI$.
Similarly, $\cN_j$ is the set of neighbors of check node $j\in\cJ$.
Denote by $\cC_j$ the constituent binary single parity check code
corresponding to node $j\in\cJ$. Let $\bzero \in \cC$ denote the
all-zero codeword.

The LP decoder~\cite{lpdecode} solves the following optimization
problem.

\bre (\blambda,\blambda_{\bomega}) \defined
\argmin_{(\bc,\footnotesize{\bomega})} \P(\bc) \label{eq: lp
problem} \ere (i.e., $\blambda$ is the optimal $\bc$-vector and
$\blambda_{\bomega}$ is the optimal $\bomega$-vector) subject to
\bre w_{j,\bg} \ge 0 \qquad \forall j \in \cJ \: , \: \bg \in \cC_j
\label{eq:positive_w} \ere \bre \sum_{\bg \in \cC_j} w_{j,\bg} = 1
\qquad \forall j \in \cJ \label{eq:w_sum_1} \ere \bre c_i =
\sum_{\bg\in \cC_j \: , \: g_i=1} w_{j,\bg} \qquad \forall j\in\cJ
\: , \: i \in \cN_j \label{eq:ci} \ere where the vector $\bomega$ is
defined by \bre \label{eq:w_def} \bomega
\defined \{ w_{j,\bg} \}_{j\in\cJ,\: \bg \in \cC_j} \nonumber \ere
and where \bre \P(\bc) \defined \sum_{i\in\cI} c_i \gamma_i
\label{eq:mllp} \ere and \bre \gamma_i
\defined \log \frac{Q(y_i \given 0)}{Q(y_i \given 1)}
\label{eq:gamma_def} \nonumber \ere is the log-likelihood ratio
(LLR). All logarithms are natural.

An important observation is that the LP decoder has the ML
certificate~\cite{lpdecode} in the sense that if the solution
$\blambda$ is integer (in fact an integer $\blambda_{\bomega}$
implies that $\blambda$ is also integer by~\eqref{eq:ci}) then it is
the ML decision.

\section{The approximate ML certificate property} \label{sec:approx_ML_cert}
Let the binary LDPC code $\cC$ be selected uniformly from an
ensemble of LDPC codes $\cC^0$ (e.g., the ensemble of all
$(c,d)$-regular codes) and assume $\cC$ has $M$ codewords. Let
$\bc_m \in \cC, m \in \{1,\dots,M\}$ be the codeword selected for
transmission over a MBIOS channel, and suppose that the channel
output vector is $\by$. The received vector $\by$ is decoded by a
standard belief propagation (BP) decoder, which outputs an estimate
$\bhc$ which may
or may not be a valid codeword. 
Let $\P(\blambda)$ be the optimal value
of the LP
problem. 
By the fact that $\bc_m$ is a codeword and hence feasible in the LP
we have \bre \label{eq:D_P_ineq} \\P(\blambda) \le \P(\bc_m) \ere
Now suppose that for some $\delta > 0$ \bre \label{eq:delta_def}
\P(\hbc) - \P(\blambda) \le \delta \ere and that the decoder output
$\bhc$ is a valid codeword. We call this event the \emph{approximate
ML certificate} and the constant $\delta$ the \emph{proximity gap}.
Formally, the AMLC happens if and only if \bre \bhc \in
\text{AMLC}(\delta) \label{eq: bhc in AMLC} \ere where \bre
\text{AMLC}(\delta) = \{ \bhc \; : \; \bhc \in \cC \;, \; \P(\hbc) -
\P(\blambda) \le \delta \} \label{eq: AMLC def} \ere When $\delta=0$
the AMLC coincides with the standard ML certificate
of~\cite{lpdecode}, since in this case the codeword $\hbc$ is the ML
solution. By~\eqref{eq:D_P_ineq}-\eqref{eq:delta_def} we conclude
that \bre \P(\hbc) - \P(\bc_m)
\le \delta \label{eq:PcPc0} \ere Now, consider the transmission of a
code chosen at random from $\cC^0$. The transmitted codeword $\bc_m$
is also selected at random from the chosen code. If the AMLC holds,
then the word error probability given that $\bc_m$ was transmitted
can be upper bounded as

\begin{equation}
\begin{split}
&\Pr  \left( \bhc \ne \bc_m \left| \begin{array}{c} \bhc \in \text{AMLC}(\delta) \\
\bc_m \text{ trans.}\end{array}\right. \right)  \\
& \le \Pr\left( \begin{array}{c} \exists \bc \in \cC \\ \bc \ne
\bc_m \end{array} : \P(\bc) \le \P(\bc_m) +
\delta \left| \begin{array}{c} \bhc \in \text{AMLC}(\delta) \\
\bc_m \text{ trans.} \end{array}\right. \right)
\\ &\le \frac{\Pr\left( \begin{array}{c} \exists \bc \in \cC \\ \bc \ne
\bc_m \end{array} : \P(\bc) \le \P(\bc_m) + \delta \; \left|\; \bc_m
\text{ trans.}\right.\right)} {\Pr \left( \bhc \in
\text{AMLC}(\delta) \; \left| \; \bc_m \text{ trans.} \right.
\right)}
\\ & = \frac{\Pr\left( \begin{array}{c} \exists \bc \in \cC \\ \bc \ne
\bc_m \end{array} \: : \: \log \left( \frac{Q(\by \given
\bc_m)}{Q(\by \given \bc)}\right) \le \delta \;|\; \bc_m \text{
trans.} \right)} {\Pr \left( \bhc \in \text{AMLC}(\delta) \;
\left|\;
 \bc_m \text{ trans.} \right.\right)}
\end{split}\label{eq:approx_ML_ineq1}
\end{equation}
where in the first inequality we used~\eqref{eq:PcPc0}, and in the
equality we used the fact that $\P(\bc) = \log
\frac{Q(\by\given\bzero)}{Q(\by \given \bc)}$ and $\P(\bc_m) = \log
\frac{Q(\by\given\bzero)}{Q(\by \given \bc_m)}$. 
Note that in \eqref{eq:approx_ML_ineq1}, the numerator depends on
the channel probability transition function and the code chosen,
while the denominator depends on the channel, the code, and also on
the decoding algorithm. We can further upper bound the expression
\eqref{eq:approx_ML_ineq1} by upper-bounding the numerator and
lower-bounding the denominator. In the process of doing so, we
eliminate the dependence on the transmitted message $m$.

\subsection{A lower bound on the denominator}
\label{sec: denominator} Consider the denominator in
\eqref{eq:approx_ML_ineq1}. The following lemma asserts its
independence of the transmitted codeword.
\begin{lemma}\label{lemma: denominator independence}
Consider the vector $\blambda$ which is output by the LP decoder.
Also assume that $\bc_m$ is the transmitted codeword and that the BP
decoder is used. Then the expression \bre \Pr \left( \bhc \in
\text{AMLC}(\delta) \;\given\; \bc_m \text{ trans.}\right)
\label{eq: bound given zero} \ere
\end{lemma}
is independent of $m$, and in particular it can be assumed in
\eqref{eq: bound given zero} that the all-zero codeword is
transmitted.
\begin{proof}
See Appendix~\ref{sec: proof of lemma1}. \end{proof} To get a lower
bound on \eqref{eq: bound given zero}, one could run Monte Carlo
simulations. Consider a series of experiments conducted to estimate
$\eta
\defined \Pr \left(
\bhc \in \text{AMLC}(\delta) \;\given\; \bzero \text{
trans.}\right)$. In each experiment we draw a code at random from
$\cC^0$, transmit the all-zero codeword over the noisy channel and
decode. Suppose that we run $L$ experiments and find that $\bhc \in
\text{AMLC}(\delta)$ in $L_1$ experiments. Let
$$ L_1 = \sum_{i=1}^L X_i
$$
where if in the $i$'th experiment $\bhc \in \text{AMLC}(\delta)$,
then $X_i = 1$; otherwise
$X_i=0$. 
If the channel is low-noise then we would expect
to have $L_1 = L(1-\epsilon)$ with small $\epsilon$, and in
particular we would expect to have $\epsilon < 0.5$, which (for
large $L$) would imply $\eta > 0.5$. Since $\eta$ is a deterministic
but unknown parameter, we cannot claim that $\eta
> 0.5$, even if $\epsilon$ is small; rather, this situation falls under the framework of
non-bayesian hypothesis testing, so the series of experiments does
allow us to say something about $\eta$ with some degree of
confidence. Consider the hypothesis \bre H_0\;:\; \eta \leq 0.5
\label{eq: hypothesis} \ere For $\epsilon<0.5$, the following
inequality holds \bre \label{eq: hypothesis test}
 \Pr(L_1 \ge (1-\epsilon) L \given H_0 \text{ valid}) <
2^{-L}\binom{L}{\epsilon L}\cdot (\epsilon L+1) \ere since the RHS
is an upper bound on the tail of a binomial distribution. Now
suppose that in a Monte Carlo simulation we get $L_1=L(1-\epsilon)$
with $\epsilon<0.5$. By \eqref{eq: hypothesis test} we observe that
if $\epsilon$ is very small, then the RHS of \eqref{eq: hypothesis
test} is very low, and thus we can reject $H_0$ with a high degree
of confidence. Conversely, if in our simulation $\epsilon$ is close
to $0.5$, we cannot reject $H_0$ with high confidence.



Define \bre \xi(L,\epsilon)\defined 1-2^{-L}\binom{L}{\epsilon
L}\cdot (\epsilon L+1) \label{eq: xi(L,eps) def}\ere Given the Monte
Carlo result discussed above, one may conclude that \bre
\begin{split} \Pr & \left( \bhc \ne \bc_m \given \bhc \in \text{AMLC}(\delta) \;,\; \bc_m \text{ trans.}\right) \\ & \le 2 \Pr\left(
\begin{array}{c} \exists \bc \in \cC \\ \bc \ne \bc_m \end{array}\: : \: \log \left(
\frac{Q(\by \given \bc_m)}{Q(\by \given \bc)}\right) \le \delta
\;\given\; \bc_m \text{ trans.}\right)\end{split} \label{eq: lb on
denominator} \ere which reflects the assertion $\eta>0.5$. This
assertion holds with confidence level $\xi(L,\epsilon)$. Note that
for fixed $\epsilon$, the likelihood that the bound \eqref{eq: lb on
denominator} does not hold decays exponentially with $L$.

The standard approach to estimating the frame error rate performance
is to use a Monte Carlo simulation. The result is a confidence
interval on the actual error rate. In our method we also use a Monte
Carlo simulation. However, in the following we derive an analytic
bound on the RHS of \eqref{eq: lb on denominator} which, combined
with the simulation, enables us to obtain extremely large confidence
levels for very small frame error rates whenever the AMLC holds.

%

\subsection{An upper bound on the numerator}
Consider now the RHS of \eqref{eq: lb on denominator} (disregarding
the constant $2$). Recalling that $\cC$ is chosen at random from
$\cC^0$, one may write
\begin{equation} \begin{split}\Pr&\left( \begin{array}{c} \exists \bc \in \cC \\ \bc \ne \bc_m \end{array} \: : \: \frac{Q(\by \given \bc)}{Q(\by \given \bc_m)}
e^{\delta} \ge 1 \;\given\; \bc_m \text{ trans.}\right) \\&=
\sum_{\cC_i\in \cC^0} \Pr\left( \begin{array}{c} \exists \bc \in \cC
\\ \bc \ne \bc_m \end{array} \: : \: \frac{Q(\by \given \bc)}{Q(\by
\given \bc_m)} e^{\delta} \ge 1 \left|\begin{array}{c} \cC=\cC_i \\
\bc_m \text{ trans.} \end{array} \right. \right) \\ & \hspace*{2cm}
\cdot \Pr\left( \cC=\cC_i \;\given\; \bc_m \text{
trans.}\right)\end{split} \label{eq: ensemble average}
\end{equation}
Clearly, $\Pr\left( \cC=\cC_i \;\given\; \bc_m \text{
trans.}\right)=\Pr\left( \cC=\cC_i\right)$ as the selection of the
message is independent of the selection of the code. In addition, we
have the following result regarding the independence of the inner
expression in the sum \eqref{eq: ensemble average} on $m$.
\begin{lemma}
The expression \bre \Pr\left(
\begin{array}{c} \exists \bc \in \cC
\\ \bc \ne \bc_m \end{array} \; : \;
\frac{Q(\by \given \bc)}{Q(\by \given \bc_m)} e^{\delta} \ge 1
\left| \begin{array}{c} \cC=\cC_i \\ \bc_m \text{ trans.}
\end{array} \right. \right)\label{eq: expression indep of m} \ere
appearing within the sum \eqref{eq: ensemble average} is independent
of $m$. \label{lemma: indep of DS2 on m}
\end{lemma}
\begin{proof} See Appendix~\ref{sec: independence of m}.
\end{proof} Due to Lemma~\ref{lemma: indep of DS2 on m}, we can assume without loss of generality that the all-zero
codeword $\bzero$ is transmitted and rewrite \eqref{eq: ensemble
average} as
\begin{equation} \begin{split} \Pr&\left( \begin{array}{c} \exists \bc \in \cC
\\ \bc \ne \bzero \end{array} \: : \: \frac{Q(\by \given \bc)}{Q(\by \given \bzero)}
e^{\delta} \ge 1 \;\given\; \bzero \text{ trans.} \right) \\&=
\sum_{\cC_i\in \cC^0} \Pr\left( \begin{array}{c} \exists \bc \in \cC
\\ \bc \ne \bzero \end{array} \: : \:
\frac{Q(\by \given \bc)}{Q(\by \given \bzero)} e^{\delta} \ge 1
\left|
\begin{array}{c} \cC=\cC_i \\ \bzero \text{ trans.} \end{array}\right.\right) \\ & \hspace*{2cm} \cdot \Pr\left(
\cC=\cC_i \right) \end{split} \label{eq: ensemble average given 0}
\end{equation}
Sason and Shamai \cite{shamai2002vgb} have proposed a tight upper
bound on the ML decoding error probability using the generalized
second version of the Duman-Salehi bound, referred to as the DS2
bound. Using a slightly modified version of this bound, we can find
an upper bound on $$\Pr\left( \begin{array}{c} \exists \bc \in \cC
\\ \bc \ne \bzero \end{array} \: : \: \frac{Q(\by
\given \bc)}{Q(\by \given \bzero)} e^{\delta} \ge 1 \left|
\begin{array}{c} \cC=\cC_i \\ \bzero \text{ trans.}\end{array}\right. \right)$$ To this end, one may write \bre
\begin{split} \Pr& \left( \begin{array}{c} \exists \bc \in \cC
\\ \bc \ne \bzero \end{array} \: : \: \frac{Q(\by \given
\bc)}{Q(\by \given \bzero)} e^{\delta} \ge 1 \left|
\begin{array}{c} \cC=\cC_i \\ \bzero \text{ trans.}\end{array}\right. \right)\\ &\le
\sum_{\by}Q(\by|\bzero) \left(\sum_{\substack{\bc \ne \bzero \\ \bc
\in \cC_i}} \left(\frac{Q(\by|\bc)}{Q(\by|\bzero)}e^{\delta}
\right)^{\lambda} \right)^{\rho} \\&=
\sum_{\by}\Psi_N^0(\by)\Psi_N^0(\by)^{-1}Q(\by|\bzero)
\left(\sum_{\substack{\bc \ne \bzero \\ \bc \in \cC_i}}
\left(\frac{Q(\by|\bc)}{Q(\by|\bzero)}e^{\delta} \right)^{\lambda}
\right)^{\rho} \\ &=
\sum_{\by}\Psi_N^0(\by)\left(\Psi_N^0(\by)^{-\frac{1}{\rho}}Q(\by|\bzero)^{\frac{1}{\rho}}
\sum_{\substack{\bc \ne \bzero \\ \bc \in \cC_i}}
\left(\frac{Q(\by|\bc)}{Q(\by|\bzero)}e^{\delta} \right)^{\lambda}
\right)^{\rho}\end{split} \label{eq: DS2 b4 Jensen}\ere where the
expression on the second line, which holds for all $\lambda,\rho \ge
0$, is an adaptation of the 1965 Gallager bound \cite{galbook} to
our purposes, and $\Psi_N^0(\by)$ is a probability measure on $\by$
called a \emph{tilting measure} \cite{shamai2002vgb}, which is
allowed in general to depend on the transmitted codeword. By
invoking Jensen's inequality in \eqref{eq: DS2 b4 Jensen}, we get
\bre
\begin{split}\Pr&\left(
\begin{array}{c} \exists \bc \in \cC
\\ \bc \ne \bzero \end{array} \: : \: \frac{Q(\by \given \bc)}{Q(\by \given
\bzero)} e^{\delta} \ge 1 \left| \begin{array}{c} \cC=\cC_i \\
\bzero \text{ trans.}\end{array} \right. \right) \\ &\le
\left(\sum_{\substack{\bc \ne \bzero
\\ \bc \in \cC_i}} \sum_{\by}
Q(\by|\bzero)^{\frac{1}{\rho}}\Psi_N^0(\by)^{1-\frac{1}{\rho}}
\left(\frac{Q(\by|\bc)}{Q(\by|\bzero)}e^{\delta} \right)^{\lambda}
\right)^{\rho}\end{split} \label{eq: DS2 vector}\ere which holds for
$\lambda\ge 0, \; 0 \le \rho \le 1$. Now let us restrict our
discussion to tilting measures which do not depend on the
transmitted codeword and which also decompose as $N$-fold products
of the same single-letter measure, i.e., \bre \Psi_N^0(\by) =
\prod_{i=1}^N \psi(y_i) \nonumber \ere Also recall that the channel
is memoryless and thus also decomposes as an $N$-fold product. Using
this in \eqref{eq: DS2 vector} yields
\begin{equation} \begin{split}\Pr&\left( \begin{array}{c} \exists \bc \in \cC
\\ \bc \ne \bzero \end{array} \: : \:
\frac{Q(\by \given \bc)}{Q(\by \given \bzero)} e^{\delta} \ge 1 \left| \begin{array}{c} \cC=\cC_i \\
\bzero \text{ trans.}\end{array} \right. \right) \\
&\le e^{\delta\rho\lambda} \left[\sum_{h=1}^N A_h \left(\sum_y
\psi(y)^{1-\frac{1}{\rho}}Q(y|0)^{\frac{1}{\rho}} \right)^{N-h}
\right. \\ & \left. \quad \cdot
\left(\sum_y\psi(y)^{1-\frac{1}{\rho}}
Q(y|0)^{\frac{1-\lambda\rho}{\rho}}Q(y|1)^{\lambda} \right)^{h}
\right]^{\rho} \end{split} \label{eq: DS2
singleletter}\end{equation} where $A_h$ is the distance spectrum of
the code $\cC_i$. Now we partition the code $\cC_i$ into constant
Hamming weight subcodes where $\cC_{i,h}$ contains all words in
$\cC_i$ of weight $h$ (note that in general these subcodes are
nonlinear). By applying a union bound over the subcodes on the LHS
of \eqref{eq: DS2 singleletter} we get
\begin{equation} \begin{split}\Pr&\left( \begin{array}{c} \exists \bc \in \cC
\\ \bc \ne \bzero \end{array} \:
: \: \frac{Q(\by \given \bc)}{Q(\by \given \bzero)} e^{\delta} \ge 1
| \cC=\cC_i \;,\; \bzero \text{ trans.} \right)\\ & \le \sum_{h=1}^N
\Pr\left( \exists \bc \in \cC_{i,h} \: : \: \frac{Q(\by \given
\bc)}{Q(\by \given \bzero)} e^{\delta} \ge 1 \left| \begin{array}{c} \cC=\cC_i \\
\bzero \text{ trans.}\end{array} \right. \right) \\ & \defined
\sum_{h=1}^N P_1(h)\end{split} \label{eq: union bound over subcodes}
\end{equation}
where by \eqref{eq: DS2 singleletter} \bre \begin{split} P_1(h) \le
& e^{\delta\rho\lambda} (A_h)^{\rho} \left[ \left(\sum_y
\psi(y)^{1-\frac{1}{\rho}}Q(y|0)^{\frac{1}{\rho}} \right)^{N-h}
\right. \\ & \left. \cdot \left(\sum_y\psi(y)^{1-\frac{1}{\rho}}
Q(y|0)^{\frac{1-\lambda\rho}{\rho}}Q(y|1)^{\lambda} \right)^{h}
\right]^{\rho} \end{split} \label{eq: DS2 subcode} \ere Let
$\overline{A_h}$ and $\overline{P_1(h)}$ denote the averages of the
distance distribution $A_h$ and $P_1(h)$, respectively, taken over
the ensemble $\cC^0$. By Jensen's inequality (applied as
$\overline{\left[(A_h)^{\rho} \right]}\le \left[\overline{A_h}
\right]^{\rho}, 0~\le~\rho~\le~1$) we have \bre
\begin{split} \overline{P_1(h)} \le & e^{\delta\rho\lambda}
(\overline{A_h})^{\rho} \left[ \left(\sum_y
\psi(y)^{1-\frac{1}{\rho}}Q(y|0)^{\frac{1}{\rho}} \right)^{N-h}
\right. \\ & \left. \cdot \left(\sum_y\psi(y)^{1-\frac{1}{\rho}}
Q(y|0)^{\frac{1-\lambda\rho}{\rho}}Q(y|1)^{\lambda} \right)^{h}
\right]^{\rho} \end{split} \label{eq: DS2 subcode average} \ere The
overall bound is given by \bre \Pr \left( \bhc \ne \bc_m \given \bhc
\in \text{AMLC}(\delta) \;,\; \bc_m \text{ trans.}\right) \le 2
\sum_{h=1}^N \overline{P_1(h)} \label{eq: overall DS2 bound} \ere
where $\overline{P_1(h)}$ is given by \eqref{eq: DS2 subcode
average}. This upper bound only depends on the average distance
spectrum, which is known for many code ensembles and in particular
for LDPC codes. Now, we can optimize the bound \eqref{eq: DS2
subcode average} over $\lambda \ge 0$\;,\;$0 \le \rho \le 1$ and the
tilting measure $\psi(\cdot)$. This optimization is performed for
every value of $h$ \emph{separately}. Some additional technical
details regarding this optimization are provided in
Appendix~\ref{sec: DS2 optimization}.

\subsection{Application of the AMLC to expurgated LDPC ensembles}
In this subsection we consider the application of the upper bound on
the error probability given the AMLC to expurgated ensembles of LDPC
codes. The expurgated ensemble $\cC^{\gamma}$ is obtained from the
original ensemble $\cC^0$ by removing all codes with minimum
distance $\gamma$ or less. The reason for dealing with this ensemble
rather than $\cC^0$ is that the decoding error probability over
$\cC^0$ is dominated \cite{gallager_LDPC_article,asympt_enum} by a
small set of ``bad'' codes with small minimum distance; we will show
that if we can avoid these ``bad'' codes, then the occurrence of the
AMLC implies very low error rates.

Let $\overline{A_h^{\gamma}}$ denote the average distance spectrum
over $\cC^{\gamma}$. It was shown \cite{asympt_enum} that if
$\gamma>0$ is selected small enough, then with probability $1-o(1)$,
a randomly-selected code from $\cC^0$ is also in $\cC^{\gamma}$;
this implies that for large enough $N$, so that less than half the
codes are expurgated, the following bound holds: \bre
\overline{A_h^{\gamma}} \leq \left\{
\begin{array}{l l} 2\overline{A_h}, & h>\gamma \\ 0, & h \le \gamma \end{array} \right. \label{eq: expurgated Ah} \ere
When using the DS2 bound we can plug $\overline{A_h^{\gamma}}$
instead of $\overline{A_h}$ in \eqref{eq: DS2 subcode average}. In
practice, when applying the Monte Carlo procedure outlined in
Section~\ref{sec: denominator}, we draw codes at random from $\cC^0$
and thus we need to test whether these codes are also in
$\cC^{\gamma}$. To do this, we use the procedure described in
\cite[Section 5]{burshtein2010mllp} which obtains a lower bound
$LB(\cC_1)$ on the minimum distance of the randomly-drawn code
$\cC_1$. If $LB(\cC_1)>\gamma$, then $\cC_1 \in \cC^{\gamma}$. Note,
however, that the converse is not necessarily true, i.e., we could
have $\cC_1 \in \cC^{\gamma}$ but with $LB(\cC_1)\le\gamma$. Define
the ensemble \bre \tilde{\cC}^{\gamma} = \{ \cC \in \cC^0\;:\;
LB(\cC)>\gamma\} \ere then clearly $\tilde{\cC}^{\gamma} \subseteq
\cC^{\gamma}$. Let $\overline{\tilde{A}_h^{\gamma}}$ be the average
distance spectrum over $\tilde{\cC}^{\gamma}$. We will obtain an
upper bound on $\overline{\tilde{A}_h^{\gamma}}$ which is similar to
\eqref{eq: expurgated Ah} using a Monte Carlo approach, similar to
the argument made in Section~\ref{sec: denominator}. Suppose we run
$L$ experiments. In each experiment we randomly pick a code $\cC \in
\cC^0$ and calculate $LB(\cC)$. Now suppose that in
$L_2=L(1-\epsilon_2)$ experiments we obtain that $LB(\cC)>\gamma$,
and $\epsilon_2<0.5$ is small. From this set of experiments, we
conclude as we did in Section~\ref{sec: denominator} that \bre
\overline{\tA_h^{\gamma}} \leq \left\{
\begin{array}{l l} 2\overline{A_h}, & h>\gamma \\ 0, & h \le \gamma \end{array} \right. \label{eq: expurgated Ah with lb} \ere
with high confidence level.

Consider the following procedure for obtaining a bound on the
confidence level of \eqref{eq: DS2 subcode average}-\eqref{eq:
overall DS2 bound} when $\overline{\tA_h^{\gamma}}$ (upper-bounded
in \eqref{eq: expurgated Ah with lb}) is used as the distance
spectrum. The confidence level output by this algorithm is a
combination of the confidence level associated with $\eta \ge 0.5$
(see Section~\ref{sec: denominator}) and the statement \eqref{eq:
expurgated Ah with lb}. That is, the null hypothesis in this case is
\bre H_0\;:\; \{ \eta \le 0.5 \text{ or } \Pr(LB(\cC)>\gamma) \le
0.5\} \label{eq: null hypothesis} \ere
\begin{algorithm}\label{alg: conf level}
Given an ensemble of codes $\cC^0$, a channel probability
distribution $Q(\cdot|\cdot)$ and number of trials $L$, do:
\begin{enumerate}
\item {\bf Initialize:} Set $E=0$.
\item {\bf Loop} $L$ times:
    \begin{itemize}
    \item Pick a code $\cC$ uniformly from $\cC^0$.
    \item Calculate $LB(\cC)$.
    \item If $LB(\cC) \le \gamma$, $E\leftarrow E+1$ and skip to next loop
    iteration.
    \item Transmit the all-zero codeword through the channel.
    \item Decode using the BP decoder and the LP decoder.
    \item If the BP decoder output $\bhc$ is not a codeword, or if
    $\P (\bhc)-\P(\blambda)>\delta$, set $E\leftarrow E+1$
    \end{itemize}
\item {\bf Output confidence level of bound:} Define $\epsilon\defined E/L$.
If $\epsilon<0.5$, output $\xi(L,\epsilon)$ defined in \eqref{eq:
xi(L,eps) def}. Otherwise, output ``error''.
\end{enumerate}
\end{algorithm}

Algorithm~\ref{alg: conf level} is introduced for the purpose of
jointly assessing the possibility of rejecting the hypothesis
\eqref{eq: hypothesis}, and the validity of \eqref{eq: expurgated Ah
with lb} as an upper bound on the distance spectrum using the same
confidence level-based Monte Carlo based method from
Section~\ref{sec: denominator}. The algorithm counts the number of
failed attempts $E$ out of $L$ experiments, where a failure consists
of either having a code $\cC$ not pass the test $LB(\cC) > \gamma$,
or, having passed this test, getting a BP decoder output which does
not satisfy the AMLC. The algorithm is correct because if the null
hypothesis \eqref{eq: null hypothesis} holds, then in any single
experiment we would have a probability of failure at least $0.5$. If
the total number of failures $E$ is small (i.e., less than half the
total number of experiments) then the confidence level, following
the derivation in Section~\ref{sec: denominator}, is output. On the
other hand, if $E \ge 0.5L$, the result is deemed unreliable.

\subsection{Statement of Main Result}
The analysis in this section leads to the following result.
\begin{theorem}
Consider the transmission of a codeword from an LDPC code drawn at
random from the ensemble $\cC^0$ over an MBIOS channel. Fix the
proximity gap $\delta>0$ and the expurgation depth $\gamma>0$. Then
the probability of frame error with BP decoding given that the AMLC
\eqref{eq: bhc in AMLC}-\eqref{eq: AMLC def} holds is upper-bounded
by \eqref{eq: DS2 subcode average}-\eqref{eq: overall DS2 bound}.
This bound holds with confidence level $\xi(L,\epsilon)$ which can
be obtained using the $L$ Monte Carlo experiments, as detailed in
Algorithm~\ref{alg: conf level}.
\end{theorem}

\section{Numerical Results and Discussion} \label{sec: numerical results}
Figure~1 shows a comparison between the frame error rate (FER)
obtained by a simulation of the BP decoder over the binary symmetric
channel (BSC) and the DS2 bound \eqref{eq: overall DS2 bound},
calculated for various values of $\delta$. In this example, we
consider the ensemble $\cC^0$ of (3,4)-regular LDPC codes with block
length $N=1000$. For the calculation of the DS2 bound and the
distance spectrum we use $\gamma=20$ as the expurgation depth.

We conducted two experiments to determine the confidence level of
the bound, using Algorithm~\ref{alg: conf level}. In the first
experiment, $150$ randomly-generated codes were tested over a BSC
with crossover probability $p=0.14$. In the second experiment, $600$
randomly-generated codes were tested over a BSC with $p=0.1$. In
both experiments, all the codes belonged to the ensemble
$\cC^{\gamma}$. The results of the first experiment are summarized
in Table~\ref{table: conf level}. These results indicate that in
this case, the null hypothesis \eqref{eq: null hypothesis} can be
rejected with very high confidence level even for $\delta=0$.
Consequently, the conditional frame error probability given that the
the AMLC holds for $\delta=0$, is (with very high confidence) lower
than $3\cdot 10^{-5}$, which is about $1000$ times lower than the
simulated frame error rate at $p=0.14$.
\begin{figure}
\begin{center}
\includegraphics[scale=0.6]{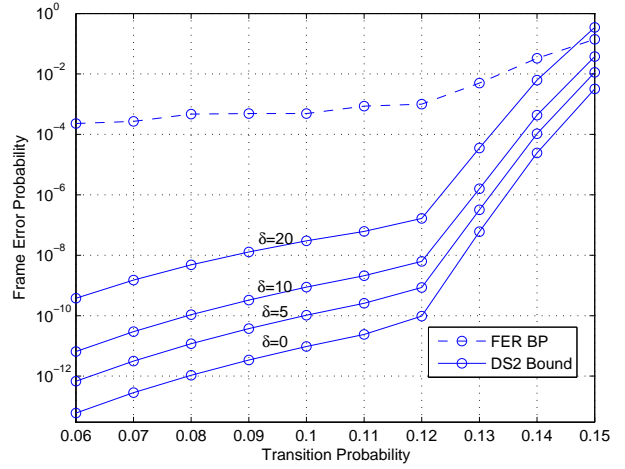}\label{fig: results1}
\caption{A comparison between simulated frame error rate of the BP
decoder and the DS2 bound \eqref{eq: overall DS2 bound}, assuming
the AMLC, for the (3,4)-regular LDPC ensemble and various values of
$\delta$.}
\end{center}
\end{figure}
In the second experiment, both the BP and LP decoders succeeded in
decoding all transmissions, and thus in Algorithm~\ref{alg: conf
level} we get $\epsilon=0$. This puts the confidence level of all
the DS2 bound curves in Figure~1 at an extremely high level of \bre
\xi(600,0)= 1-2^{-600} \label{eq: high conflevel}\ere In this case,
the conditional frame error probability given that the AMLC holds is
more than $7$ orders of magnitude smaller than the simulated frame
error rate (the difference between the BP curve and the $\delta=0$
curve for $p=0.1$). The confidence level in this case is also much
higher than in the first experiment. Due to the high confidence
levels observed in the first experiment with $p=0.14$, the
confidence level result of the second experiment with $p=0.1$ is not
surprising, and in general we expect the confidence level to
increase as the channel noise level decreases.
\begin{table}
\begin{center}
\begin{tabular}{|c|c|c|}
\hline $\delta$ & $\epsilon$ & $1-\xi(L,\epsilon)$ \\ \hline
$0$ & $0.1667$ & $7.13\cdot 10^{-31}$ \\
$5$ & $0.0733$ & $4.95\cdot 10^{-37}$ \\
$10$ & $0.02$ & $1.6\cdot 10^{-42}$ \\
$20$ & $0$ & $7\cdot 10^{-46}$\\ \hline
\end{tabular}
\end{center}
\caption{Confidence level bounds for different values of $\delta$}
\label{table: conf level}
\end{table}

The strength of this result is that it demonstrates that the LP
decoder can provide the BP decoder with extra error detection
capability. This capability is especially useful in applications
where a codeword should be rejected unless it is decoded correctly,
and rejection should occur with high probability (as in data
applications requiring high reliability). Achieving codeword
reliability results of this order via simple Monte Carlo simulation
would require very long simulation runs. In fact, our technique for
upper bounding the frame error rate given that the AMLC property
holds, has a common feature with the importance sampling method,
since both attempt to alleviate the computational burden associated
with simple Monte Carlo simulation. We also note that in both
experiments described above, the AMLC was satisfied in a large
percentage of the trials, implying that it is not only capable of
increasing the reliability of the decoder output, but it also does
so very \emph{frequently}. Using the AMLC provides an alternative to
external error-detection codes, such as cyclic redundancy checks,
which cause some coding rate loss. This comes at the expense of
extra processing, in the shape of an LP decoder at the receiver. We
note that this decoder can be implemented in linear time
\cite{lpon_journal}. There is also a one-time task of computing the
confidence level, which can be performed off-line. The computational
complexity of calculating the lower bound \cite{burshtein2010mllp}
on the minimum distance $LB(\cC)$ is quadratic in the block length.
Thus the task of obtaining a confidence level using
Algorithm~\ref{alg: conf level} is performed with complexity
$O(N^2L)$, where $L$ is the number of simulated blocks. We also note
the following points.
\begin{itemize}
\item It is possible to tune the AMLC result to obtain
different error rates and confidence levels by varying the value of
the proximity gap $\delta$ and the expurgation depth $\gamma$.
Higher values of $\delta$ will produce higher values of the DS2
bound (this is evident from Figure~1), but on the other hand will
increase the confidence level (as can be seen in Table~\ref{table:
conf level}), as the requirement \eqref{eq:delta_def} becomes more
lax. Higher values of $\gamma$ will yield lower values of the DS2
bound. This, however, comes at the expense of a lower confidence
level because while running Algorithm~\ref{alg: conf level} more
codes will be rejected as having low minimum distance.
\item
One may observe that in the example above, the AMLC result is
applied to a \emph{random selection of a code from an ensemble}. In
many applications, it would be desirable to apply the AMLC result to
a specific code rather than an ensemble. The difficulty is that
while \emph{ensemble averages} of distance spectra are typically
known or can be easily upper-bounded, for specific codes this is not
the case in general. Naturally, if one obtains for a specific code
the exact distance spectrum (or an upper bound), it is
straightforward to plug it in the DS2 bound \eqref{eq: DS2 subcode
average}-\eqref{eq: overall DS2 bound}. Another alternative is to
use known concentration results \cite{Rathi2006} for the distance
spectrum which enable one to give upper bounds on the distance
spectrum of a specific code, which themselves hold with some
confidence level. This confidence level can be integrated with our
confidence bound $\xi(L,\epsilon)$. The result would be a looser
bound (as compared to \eqref{eq: DS2 subcode average}-\eqref{eq:
overall DS2 bound}) which applies with confidence level worse than
$\xi(L,\epsilon)$, but it would apply to specific codes.
\item
Application of the AMLC result is not restricted to the BP decoder.
The result extends trivially to any decoder which satisfies the
symmetry condition \eqref{eq: BP symmetry}. In particular, this
condition is fulfilled by standard message-passing algorithms, e.g.,
min-sum, Gallager-A, Gallager-B.
\item
In a more general context, the AMLC result can be applied to any LP
formulation. In particular, the LP program proposed by Feldman
\cite{Feldman_phd} for general Turbo codes can be used to achieve
better error detection under standard iterative decoding schemes
\footnote{For standard parallel concatenated Turbo codes, no
expurgation is needed because all codes in the ensemble do not have
codewords of very low weight.}. The same goes for nonbinary LDPC
codes when represented using the LP formulation proposed by Flanagan
\etal \cite{flanagan2008lpd}. 
\end{itemize}

Finally, it may be observed that our bound can be improved by any
method which tightens the LP relaxation, e.g., the check node
merging technique \cite{burshtein2010mllp}, lifting
methods~\cite{lpdecode}, and others. By using any of these methods,
we can obtain a vector $\blambda$ such that $\P(\blambda)$ is
\emph{larger} than that obtained by the standard LP decoder,
essentially because the optimization \eqref{eq: lp problem} is
performed over a smaller domain. The result is that for any
BP-decoded codeword $\bhc$, we can use a smaller value of $\delta$
in the AMLC \eqref{eq: AMLC def}, which gives an exponential
improvement in the DS2 bound, as can be clearly seen in~\eqref{eq:
DS2 subcode average}-\eqref{eq: overall DS2 bound}.

\section{Conclusion}\label{sec: conclusion}
A new property, the approximate maximum-likelihood certificate, is
introduced. This property of a BP-decoded codeword enables to
increase its reliability, i.e., to increase the error detection
capability. This is achieved for LDPC codes using tools related to
linear programming decoding, including a recently-proposed algorithm
for finding a lower bound on the minimum distance of a specific code
which serves to improve the result. By applying the AMLC in the
error floor region, it was demonstrated that the property can imply
a frame error rate several orders of magnitude lower than a
simulated error rate. While the increased frame error detection
capability only holds with a certain confidence level, it was shown
that this level is extremely high in
the error floor region. 


\appendices

\section{Proof of Lemma \ref{lemma: denominator independence}} \label{sec: proof of lemma1}
Consider first the BP decoder. From the symmetry of the BP algorithm
over MBIOS channels \cite{urbanke_capacity}, we know that \bre
\bhc_i((-1)^{\bc_m}\cdot \by)= \left\{
\begin{array}{l l} \bhc_i(\by), & \bc_{m,i}=0 \\ 1-\bhc_i(\by), & \bc_{m,i}=1 \end{array} \right. \label{eq: BP symmetry}\ere
where $(-1)^{\bc_m}$ is a vector of $\pm 1$ corresponding to the
codeword $\bc_m$, the multiplication $(-1)^{\bc_m}\cdot \by$ is
componentwise, and $\bhc_i$ (resp. $\bc_{m,i}$) is the $i$'th bit of
$\bhc$ (resp. $\bc_m$). Now consider the LP decoder, which is used
to produce the vector $\blambda$. Fix $\epsilon>0$. The vector
$\blambda=\{\lambda_i\}_{i\in \cI}$ satisfies the symmetry condition
(see \cite{lpdecode},\cite[Lemma 6]{lpon_journal}) \bre
\lambda_i((-1)^{\bc_m}\cdot
\by)=\left\{\begin{array}{l l} \lambda_i(\by), & \bc_{m,i}=0 \\
1-\lambda_i(\by), & \bc_{m,i}=1
\end{array} \right. \label{eq: LP symmetry} \ere Now,
\begin{equation} \begin{split}
\Pr & \left( \bhc \in \text{AMLC}(\delta) \;\given\; \bc_m \text{ trans.}\right) \\
&=\Pr  \left( \P\left(\bhc\right)-\P\left(\blambda\right) \le
\delta,\; \bhc \in \cC \;\given\; \bc_m \text{ trans.}\right)
\nonumber
\\ &= \Pr \left( \sum_{i\in \cI} \log
\left(\frac{Q(y_i|0)}{Q(y_i|1)} \right) \left( \bhc_i(\by) \right.
\right.
 \\ & \hspace*{1.2cm} \left. -\lambda_i(\by)\right) \leq \delta, \; \bhc
\in \cC \;\given\; \bc_m \text{ trans.} \Biggr)
\\
& = \Pr \left( \sum_{i\in \cI} \log \left(\frac{Q(y_i|0)}{Q(y_i|1)}
\right)(-1)^{\bc_{m,i}} \left( \bhc_i((-1)^{\bc_m}\cdot\by) \right.
\right. \\ & \hspace*{0.8cm} \left.
-\lambda_i((-1)^{\bc_m}\cdot\by)\right) \leq \delta,\; \bhc \in \cC
\;\given\; \bzero
\text{ trans.} \Biggr)  \\
&= \Pr \left( \sum_{i\in \cI} \log \left(\frac{Q(y_i|0)}{Q(y_i|1)}
\right)\left( \bhc_i(\by)-\lambda_i(\by)\right) \leq \delta, \right.
\\ & \hspace*{1.5cm} \bhc \in \cC \;\given\; \bzero
\text{ trans.} \Biggr)  \\
&= \Pr \left( \P(\bhc)-\P(\blambda) \leq \delta,\; \bhc \in \cC
\;\given\; \bzero \text{ trans.} \right) \\
&= \Pr \left( \bhc \in \text{AMLC}(\delta) \;\given\; \bzero \text{
trans.} \right)
\end{split}
\end{equation}
where \begin{itemize} \item the first inequality is by the
definition \eqref{eq: AMLC def}. \item in the second equality we use
the definition \eqref{eq:mllp} and stress the dependence on $\by$.
\item in the third equality we use the symmetry of the channel as well as the symmetry
of the BP decoder \eqref{eq: BP symmetry}.
\item in the fourth equality we use the symmetry of the LP and BP decoders
(\eqref{eq: BP symmetry},\eqref{eq: LP symmetry}).
\item in the fifth equality we again use the definition \eqref{eq:mllp}.
\item the final equality is again by the definition \eqref{eq: AMLC
def}.
\end{itemize}
The above series of equalities hold for all $m$. This proves the
claim.

\section{Proof of lemma \ref{lemma: indep of DS2 on m}} \label{sec: independence of m}
For any two codewords $\bc_1$ and $\bc_2$, define the sets \bra
A_1(\bc_1,\bc_2) &\defined& \{i: (\bc_1)_i=0 \;,\; (\bc_2)_i=1\}
\nonumber \\ A_2(\bc_1,\bc_2) &\defined& \{i: (\bc_1)_i=1 \;,\;
(\bc_2)_i=0\} \label{eq: definition of A1 and A2}\era where
$(\bc)_i$ is the $i$'th bit of codeword $\bc$. We now have
\begin{equation} \begin{split} \Pr&\left( \begin{array}{c} \exists \bc \in \cC
\\ \bc \ne \bc_m \end{array} \: : \: \frac{Q(\by
\given \bc)}{Q(\by \given \bc_m)} e^{\delta} \ge 1 \left|
\begin{array}{c} \cC=\cC_i \\ \bc_m \text{ trans.}\end{array}
\right. \right)  \\ & = \Pr\left( \begin{array}{c} \exists \bc \in
\cC \\ \bc \ne \bc_m \end{array} \: : \: \prod_{i \in
A_1(\bc,\bc_m)} \frac{Q(y_i \given 0)}{Q(y_i \given 1)} \right.
\\ & \hspace*{1cm} \left.\cdot \prod_{i \in A_2(\bc,\bc_m)}
\frac{Q(y_i \given 1)}{Q(y_i \given 0)}
e^{\delta} \ge 1 \left| \begin{array}{c} \cC=\cC_i \\ \bc_m \text{ trans.}\end{array} \right. \right) \\
&= \Pr\left( \begin{array}{c} \exists \bc \in \cC
\\ \bc \ne \bc_m \end{array} \: : \: \prod_{i \in A_1(\bc,\bc_m)} \frac{Q(-y_i
\given 0)}{Q(-y_i \given 1)} \right.
\\ & \hspace*{1cm} \left. \cdot \prod_{i \in A_2(\bc,\bc_m)} \frac{Q(y_i \given 1)}{Q(y_i
\given 0)} e^{\delta} \ge 1 \left| \begin{array}{c} \cC=\cC_i \\
\bzero \text{ trans.}\end{array} \right. \right)
\\ &= \Pr\left( \begin{array}{c} \exists \bc \in \cC
\\ \bc \ne \bc_m \end{array} \: : \: \prod_{i \in
A_1(\bc,\bc_m)\cup A_2(\bc,\bc_m)} \frac{Q(y_i \given 1)}{Q(y_i
\given 0)} \right. \\ & \left. \hspace*{2cm} \cdot e^{\delta} \ge 1 \left| \begin{array}{c} \cC=\cC_i \\
\bzero \text{ trans.}\end{array} \right. \right) \\ & = \Pr\left(
\begin{array}{c} \exists \bc' \in \cC
\\ \bc' \ne \bzero \end{array}  \: : \: \frac{Q(\by
\given \bc')}{Q(\by \given \bzero)} e^{\delta} \ge 1 \left| \begin{array}{c} \cC=\cC_i \\
\bzero \text{ trans.}\end{array} \right. \right)
\end{split} \end{equation} where \begin{itemize} \item the first equality is due to the definition
\eqref{eq: definition of A1 and A2} of the sets $A_1$ and $A_2$.
\item the second equality is due to the same definitions of $A_1$
and $A_2$ as well as the symmetry of the channel
($Q(y|x)=Q(-y|1-x)$).\item the third equality is due to the symmetry
of the channel. \item the final equality, which is the desired
result, is due to the linearity of the code.
\end{itemize}
\section{Optimization of the DS2 bound} \label{sec: DS2 optimization}
Consider the DS2 bound \eqref{eq: DS2 subcode average} for fixed
$h$. Let $\beta\defined \frac{h}{N}$. First, rewrite the bound in
exponential form as \begin{equation*} \overline{P_1(h)} \le e^{-N
E_{\text{DS2}}(\delta,\beta,\rho,\lambda,\psi(\cdot))}
\end{equation*}
\begin{equation*}\begin{split}
E_{\text{DS2}}&(\delta,\beta,\rho,\lambda,\psi(\cdot)) \defined
-\frac{\delta}{N}\rho\lambda - \frac{\rho}{N} \ln
\left(\overline{A_h}\right) \\ & - \rho(1-\beta) \ln \left(\sum_y
\psi(y)^{1-\frac{1}{\rho}}Q(y|0) ^{\frac{1}{\rho}} \right) \\ & -
\rho \beta \ln \left(\sum_y\psi(y)^{1-\frac{1}{\rho}}
Q(y|0)^{\frac{1-\lambda\rho}{\rho}}Q(y|1)^{\lambda}
\right)\end{split} \label{eq: DS2 subcode average exponential}
\end{equation*} Assuming fixed values of $\beta$ and $\delta$, the
exponent $E_{\text{DS2}}(\delta,\beta,\rho,\lambda,\psi(\cdot))$
should be maximized over \bre \lambda \ge 0, \quad 0 \le \rho \le 1,
\quad \left\{\psi(y): \sum_y \psi(y)=1\right\} \ere For fixed values
of $\lambda$ and $\rho$, we use calculus of variations to find the
optimum tilting measure $\psi$; this analysis yields the optimality
condition \begin{multline}
\psi(y)^{-\frac{1}{\rho}}\left(\frac{(1-\beta)(1-\frac{1}{\rho})g_1(y)}{\sum_y
\psi(y)^{1-\frac{1}{\rho}}g_1(y)} \right.
\\ \left. +\frac{\beta(1-\frac{1}{\rho})g_2(y)}{\sum_y
\psi(y)^{1-\frac{1}{\rho}}g_2(y)} \right) + \mu =0 \label{eq: DS2
optimality cond} \end{multline} where $\mu$ is a Lagrange multiplier
and \bre g_1(y)\defined Q(y|0)^{\frac{1}{\rho}} \quad g_2(y)
\defined Q(y|0)^{\frac{1}{\rho}} \left(
\frac{Q(y|1)}{Q(y|0)}\right)^{\lambda} \nonumber \ere The solution
to \eqref{eq: DS2 optimality cond} is given in the following
implicit form \bre \psi(y)= \zeta \left(g_1(y)+ \kappa g_2(y)
\right)^{\rho}= \zeta
Q(y|0)\left[1+\kappa\left(\frac{Q(y|1)}{Q(y|0)} \right)^{\lambda}
\right]^{\rho} \nonumber \ere where \bre
\kappa=\frac{\beta}{1-\beta}\frac{\sum_y Q(y|0)\left(1+\kappa
\left(\frac{Q(y|1)}{Q(y|0)} \right)^{\lambda}
\right)^{\rho-1}}{\sum_y Q(y|0)\left(\frac{Q(y|1)}{Q(y|0)}
\right)^{\lambda}\left(1+\kappa \left(\frac{Q(y|1)}{Q(y|0)}
\right)^{\lambda} \right)^{\rho-1}} \label{eq: iteration of k} \ere
The appropriate normalizing constant $\zeta$ is given by \bre \zeta=
\left[\sum_y Q(y|0)\left(1+\kappa \left(\frac{Q(y|1)}{Q(y|0)}
\right)^ {\lambda} \right)^{\rho} \right]^{-1} \label{eq: iteration
of zeta} \ere To find the optimized tilting measure, we solve
\eqref{eq: iteration of k} numerically, and determine $\zeta$ by
\eqref{eq: iteration of zeta}. The optimal values of $\lambda$ and
$\rho$ are then found numerically.

\bibliography{bibliography}

\end{document}

%% file: shortcuts.tex
\newcommand{\rem}[1]{}

\newtheorem{lemma}{Lemma}

\newtheorem{theorem}{Theorem}

\newtheorem{algorithm}{Algorithm}

\newcommand{\bre}{\begin{equation}}
\newcommand{\ere}{\end{equation}}

\newcommand{\ee}\]
\newcommand{\bra}{\begin{eqnarray}}
\newcommand{\era}{\end{eqnarray}}
\newcommand{\bfg}{\begin{figure}[hbtp]}
\newcommand{\efg}{\end{figure}}
\newcommand{\bver}{\begin{verbatim}}
\newcommand{\ever}{\end{verbatim}}
\newcommand{\bit}{\begin{itemize}}
\newcommand{\eit}{\end{itemize}}
\newcommand{\ben}{\begin{enumerate}}
\newcommand{\een}{\end{enumerate}}

\newcommand{\blambda}{\mbox{\boldmath $\lambda$}}

\newcommand{\given}{\: | \:}

\newcommand{\bomega}{\mbox{\boldmath $\omega$}}

\newcommand{\bzero}{{\bf 0}}

\newcommand{\bg}{{\bf{g}}}

\newcommand{\bc}{{\bf c}}

\newcommand{\tA}{\tilde A}

\newcommand{\bhc}{\hat {\bf c}}

\newcommand{\baa}{\begin{eqnarray*}}
\newcommand{\eaa}{\end{eqnarray*}}

\newcommand{\by}{{\bf y}}

\newcommand{\cJ}{{\cal J}}

\newcommand{\cC}{{\mathcal{C}}}
\newcommand{\cI}{{\mathcal{I}}}

\newcommand{\cN}{{\mathcal{N}}}

\newcommand{\hbc}{\hat{\bf c}}

\def\defined{\: {\stackrel{\scriptscriptstyle \Delta}{=}} \: }

\def\argmin{\mathop{\rm argmin}}

\def\defined{\: {\stackrel{\scriptscriptstyle \Delta}{=}} \: }

\newfont{\boldlarge}{msbm10 scaled 1100}

\renewcommand{\P}{{\rm P}}

\newcommand{\comment}[1]{}

\def\etal{$et \,\,al.\,\,$}

\renewcommand{\thesection}{\Roman{section}}

\newlength{\tmpbigbar}
